\documentclass[aip,jap,reprint,graphicx]{revtex4-1} % for checking your page length
\usepackage{graphicx,amssymb,amsmath,color,psfrag}
\usepackage{amsthm}
\usepackage{amsfonts}
\usepackage{algorithmic}
\usepackage{enumerate}
\usepackage{latexsym}

\begin{document}
\newcommand{\dt}{\Delta\tau}
\newcommand{\al}{\alpha}
\newcommand{\ep}{\varepsilon}
\newcommand{\ave}[1]{\langle #1\rangle}
\newcommand{\have}[1]{\langle #1\rangle_{\{s\}}}
\newcommand{\bave}[1]{\big\langle #1\big\rangle}
\newcommand{\Bave}[1]{\Big\langle #1\Big\rangle}
\newcommand{\dave}[1]{\langle\langle #1\rangle\rangle}
\newcommand{\bigdave}[1]{\big\langle\big\langle #1\big\rangle\big\rangle}
\newcommand{\Bigdave}[1]{\Big\langle\Big\langle #1\Big\rangle\Big\rangle}
\newcommand{\braket}[2]{\langle #1|#2\rangle}
\newcommand{\up}{\uparrow}
\newcommand{\dn}{\downarrow}
\newcommand{\bb}{\mathsf{B}}
\newcommand{\ctr}{{\text{\Large${\mathcal T}r$}}}
\newcommand{\sctr}{{\mathcal{T}}\!r \,}
\newcommand{\btr}{\underset{\{s\}}{\text{\Large\rm Tr}}}
\newcommand{\lvec}[1]{\mathbf{#1}}
\newcommand{\gt}{\tilde{g}}
\newcommand{\ggt}{\tilde{G}}
\newcommand{\jpsj}{J.\ Phys.\ Soc.\ Japan\ }

\title{Indirect exchange of magnetic impurities in zigzag graphene ribbon }
\author{J. H. Sun$,^{1}$ F. M. Hu$,^2$ H. K. Tang$,^{3}$ W. Guo$,^{1}$ and H. Q. Lin$^{3}$}
\affiliation{$^{1}$ Department of Physics, Peking University, Beijing 100871, China\\
$^2$COMP/Department of Applied Physics, Aalto University School of Science, P.O. Box 11100, FI-00076 Aalto, Espoo, Finland\\
$^3$Department of Physics and ITP, The Chinese University of Hong Kong, Hong Kong, China}

\begin{abstract}
We use quantum Monte Carlo method to study the indirect coupling between two magnetic impurities on the zigzag edge of graphene ribbon, with respect to the chemical potential $\mu$.
We find that the spin-spin correlation between two adatoms located on the nearest sites in the zigzag edge are drastically suppressed around the zero-energy. As we switch the system away from half-filling, the antiferromagnetic correlation is first enhanced and then decreased. If the two adatoms are adsorbed on the sites belonging to the same sublattice, we find similar behavior of spin-spin correlation except for a crossover from ferromagnetic to antiferromagentic correlation in the vicinity of zero-energy. We also calculated the weight of different components of d-electron wave function and local magnet moment for various values of parameters,
and all the results are consistent with those of spin-spin correlation between two magnetic impurities.
\end{abstract}

\date{\today}
\maketitle

\section{Introduction}
Zigzag graphene ribbon has attracted enormous attention due to its potential application in nanoscale spintronics devices.\cite{Son06,Brey07, Tombros07}
It has zero-energy localized states at Fermi level, while those with armchair edges have no such peculiar states \cite{Fujita96,Nakada96}.

Due to the edge states in zigzag graphene ribbon, the magnetic property of the Anderson impurities would be greatly modified and shows novel behavior differing from that in bulk graphene and in normal metals. When two magnetic impurities are located sufficiently far away from each other, they can be modeled by a single impurity Hamiltonian\cite{Anderson61, kondo64}. However, if they are close together, the "spin compensation clouds" of surrounding conduction electrons will overlap, and interesting effects show up due to the indirect Ruderman-Kitter-Kasuya-Yosida (RKKY) coupling \cite{RKKY1,RKKY2,RKKY3} between magnetic impurities, which is mediated by the conduction electrons.

In the bipartite lattice, the oscillatory RKKY interaction becomes commensurate and the effective exchange coupling is always ferromagnetic (FM) for the same sublattice but antiferromagnetic (AFM) for opposite sublattice at the Fermi level\cite{Saremi07,Bunder09}.
RKKY interaction on the graphene and graphenelike systems at half-filling has been constantly studied\cite{Saremi07,Bunder09}, and the RKKY coupling on zigzag edge is reported to decay exponentially due to the localized zero-energy states \cite{Annica10a}, while in graphene the coupling is proportional to $1/R^3$.\cite{Sherafait11}

In this paper, we use the quantum Monte Carlo (QMC) method based on the Hirsch-Fye algorithm \cite{Hirsch86} to study the indirect coupling between magnetic adatoms placed on the zigzag graphene edge. We calculate various thermodynamic quantities with respect to the chemical potential, since graphene is a system whose carrier density can be tuned by the electric field effect\cite{Novoselov04}.
For $T_k<T\ll D$, where $T_k$ is the Kondo temperature and D is the band width, the spin-spin correlation between the two magnetic impurities are expected to be \cite{Hirsch1987}
\begin{equation}
\begin{aligned}
\langle \sigma_{z1}\sigma_{z_2} \rangle=\frac{1-e^{\beta J(R)}}{3+e^{\beta J(R)}}\approx \frac{\beta J(R)}{4},
\end{aligned}
\end{equation}
where $\sigma_{zi}$ is the spin on the i-th impurity and $\beta$ is the inverse temperature.
Therefore our study on the spin-spin correlation between magnetic impurities can provide insights into the behavior of the RKKY interactions while $\mu \neq 0 $.

\section{Results}\label{sec:zgnrresults}
\begin{figure}[t]
\begin{center}
\includegraphics[scale=0.33, bb=220 70 560 530]{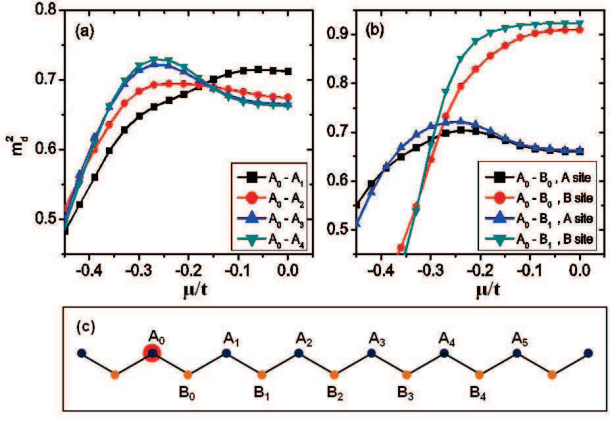}
\end{center}
\caption{(Color online). The local moment squared $m_d^2$ as a function of $\mu$. (a) Two impurities both located on the edge sites belong to sublattice A, (b) two impurities on different sublattices. $A_0-A_1$ denotes the configuration that two impurities are located on $A_0$ and $A_1$ sites.  Here we use $U=0.8t$, $V=0.5t$, $\beta=32t^{-1}$ and $\epsilon_d=-U/2$, (c) the schematic structure of zigzag chain. One adatom is fixed on the red circled site $A_0$ and the other one is moved along the zigzag chain. }\label{Fig:zgnrmd2}
\end{figure}
In our study, we choose zigzag edged ribbon with $10$ zigzag chains, where the density of states on the edge is comparable to the van Hove singularity at $E=1t$ in bulk graphene\cite{Nakada96}.

We fix the location of one impurity and change the location of the other impurity along the zigzag chain to study the behavior of the indirect interactions.
%\subsection{Magnet moment}
Shown in Fig. \ref{Fig:zgnrmd2} are the values of magnet moment squared, which is defined as
\begin{equation}
\begin{aligned}
m_d^2=\langle (n_{d\uparrow}-n_{d\downarrow})^2 \rangle,
\end{aligned}
\end{equation}
with $n_{d\sigma}$ being the carrier density on the impurity site and $\sigma$ being the spin index.
When the two impurities are both located on the edge sites belonging to sublattice A, as shown in Fig. \ref{Fig:zgnrmd2} (a), the local moment is strongly suppressed near the zero energy. Due to the indirect exchange between impurities, when two impurities are located on $A_0$ and $A_1$ sites, as shown in Fig. \ref{Fig:zgnrmd2} (c), the values of local moment is less suppressed. As the displacement between the two impurities is increased, the effect of indirect coupling decreases such that the local moment is more strongly quenched.
The results for two impurities located on two different sublattices are shown in Fig. \ref{Fig:zgnrmd2} (b). The effect of the indirect coupling is much weaker than the cases shown in Fig. \ref{Fig:zgnrmd2} (a), and we can observe that the local moment on the A sites are not affected by the RKKY interaction, while that on the B site is slightly suppressed near the zero energy. The results for site configurations with larger distance are not shown because the effect of the indirect coupling decays so fast that the results would be mostly the same.

\begin{figure}[t]
\begin{center}
\includegraphics[scale=0.52, bb=80 200 490 400]{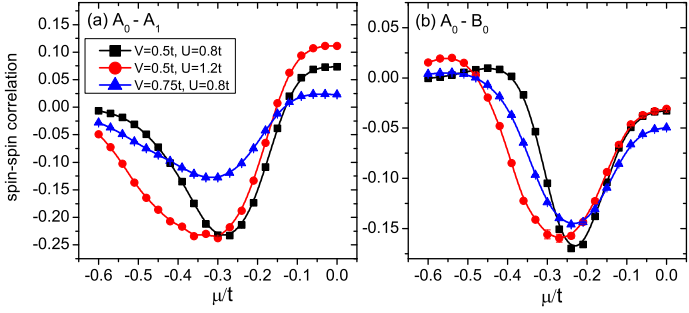}
\end{center}
\caption{(Color online). The spin-spin correlation between two magnetic impurities on the zigzag edge as a function of $\mu$. One impurity fixed on the site $A_0$,  the other impurity locatedon the site $A_1$ (a), and $B_0$ (b), respectively. Here we use $\beta=32t^{-1}$ and $\epsilon_d=-U/2$. }\label{Fig:zgnrss}
\end{figure}

%\subsection{Spin-spin correlation}
Shown in Fig.\ref{Fig:zgnrss} are the spin-spin correlation between two magnetic impurities with one impurity fixed on the site $A_0$ and the other impurity located on the site $A_1$, $B_0$, respectively. Various values of parameters as $U=0.8t,1.2t$ and $V=0.5t,0.75t$ are tested for $\beta=32t^{-1}$ and $\epsilon_d=-U/2$.
 In Fig.\ref{Fig:zgnrss} (a), we can find that when $\mu=0$, the spin-spin correlation is ferromagnetic, which is consistent with the results of the RKKY interaction in the bipartite lattice\cite{Saremi07,Bunder09} for the undoped case. As $\mu$ is lowered from zero-energy, the FM spin-spin correlation is first decreased and then at approximately at $\mu=-0.15t$, it vanishes for all the parameters we examined. If we continue to decrease the chemical potential, the FM correlation turns to AMF and the magnitude grows as $\mu$ is decreased further. Hence, we can observe a transition from FM to AFM correlation, and the AFM correlation becomes stronger when we switch down the chemical potential. The AFM correlation is the strongest at $\mu\sim-0.3t$, and the decrease in the value of $\mu$ will lead to decrease of the strength of AFM correlation. Finally at $\mu\sim-0.6t$, AFM correlation vanishes. Near the zero-energy, the spin-spin correlations for $U=0.8t$ and $V=0.75t$ is more strongly quenched compared with the case $U=0.8t$ and $V=0.5t$. As the hybridization $V$ increases, the screening effect of the conduction electron would be enhanced, which will lead to much stronger suppression on the local moment formation. On the other hand, for the same values of hybridization $V=0.5t$, the spin-spin correlation for $U=1.2t$ is larger than that of $U=0.8t$, because larger repulsive $U$ would give rise to single occupation on the impurity site.

If the second impurity is located on the site $B_0$, as is shown in Fig. \ref{Fig:zgnrss}(b), the spin-spin correlation at undoped ribbon is antiferromagnetic. However, as we switch $\mu$ below the zero-energy, the AFM correlation is enhanced, and it has a maximum magnitude at $\mu \sim -0.25t$. If we further decrease the chemical potential, the AFM correlation is decreased either, and finally at $\mu \sim -0.4t$, it vanishes and the RKKY coupling changes to a FM one.

%\subsection{Weight of different components of the d-electron wave function}
\begin{figure}[t]
\begin{center}
\includegraphics[scale=0.35, bb=100 120 600 400]{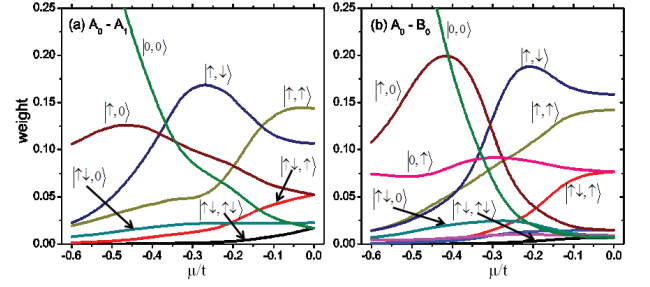}
\end{center}
\caption{(Color online). The weight of the different components of the wave function of d electrons on $A_0$ and $A_1$. Here we use $U=0.8t$, $V=0.5t$, $\beta=32t^{-1}$ and $\epsilon_d=-U/2$. }\label{Fig:zgnrweight}
\end{figure}
The QMC method naturally returns the imaginary time Green's functions of the impurity sites, by which we can simply obtain the correlation functions such as $\langle n_{i\sigma}\rangle$, $\langle n_{i\sigma}n_{j\sigma}\rangle$, $\langle n_{i\sigma}n_{i\sigma}'n_{j\sigma}\rangle$ and $\langle n_{i\sigma}n_{i\sigma}'n_{j\sigma}n_{j\sigma}'\rangle$.
The two impurity sites have $16$ different spin configurations, and the sum of the weight of all these configurations is defined to be unity. All of the configurations would contribute to the spin-spin correlation, but our simulation reveals that the configurations $|\uparrow,\downarrow\rangle$ and $|\uparrow,\uparrow\rangle$ are dominant near the zero-energy.

In Fig. \ref{Fig:zgnrweight}, we show the weight of different components of the d-electron wave function.
As is shown in Fig. \ref{Fig:zgnrweight} (a), the weight of $|\uparrow,\uparrow\rangle$ type dominates when $\mu=0$. As the chemical potential is lowered, we find that the weight of $|\uparrow,\uparrow\rangle$ type decreases and that of $|\uparrow,\downarrow\rangle$ is largely enhanced. If we further lower the chemical potential, the contribution from the $|\uparrow,0\rangle$ and the vacuum states increases due to the decrease in charge density.
In Fig. \ref{Fig:zgnrweight} (b), we notice the similar behavior. As $\mu$ is lowered, the weight of the $|\uparrow,\uparrow\rangle$ type wave function is drastically enhanced, while those of the $|\uparrow,\downarrow\rangle$ is decreased, and the two curves cross over each other at $\mu\approx-0.4t$. Hence, the studies on the weight of different components of the wave function are in agreement with those of spin-spin correlations.
As we switch down the chemical potential, parallel spin configuration becomes more and more favored by the two impurities. As $\mu$ is further lowered, the charges move out from the impurity such that the wave functions for the single occupancy and vacuum states dominate and hence, the AFM correlation is gradually decreased.

%\subsection{Comparison : two magnetic impurities in bulk graphene}
\begin{figure}[t]
\begin{center}
\includegraphics[scale=0.5, bb=20 180 500 400]{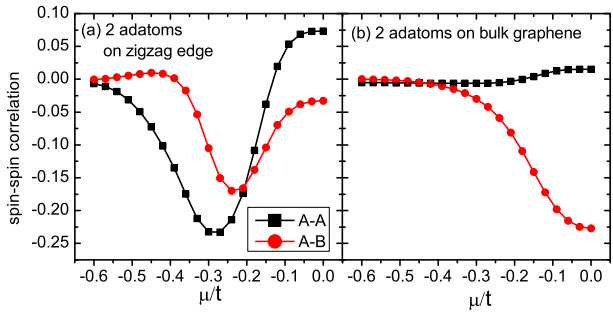}
\end{center}
\caption{(Color online). The spin-spin correlation of the two magnetic impurities located on A-A and A-B sites in bulk graphene. We fix the location of one impurity on a site belongs to the sublattice A. A-A case means the other impurity is on the next nearest neighbor(belongs to sublattice A), and A-B denotes the other impurity is located on the nearest neighbor(belongs to sublattice B). Here we use $U=0.8t$, $V=0.5t$, $\beta=32t^{-1}$. (a) Two adatoms are located on the edge of ZGNR, (b) two adatoms are located on the bulk graphene. }\label{Fig:zgnrgrapheness}
\end{figure}

Shown in Fig. \ref{Fig:zgnrgrapheness} are the comparison of spin-spin correlation between two magnetic impurities located in the zigzag ribbon and in the bulk graphene system.
In undoped pure graphene as shown in Fig. \ref{Fig:zgnrgrapheness} (b), when the two adatoms are located on the same sublattice, we can still see the FM correlation, but the strength is much smaller than that in Fig. \ref{Fig:zgnrgrapheness} (a). As we lower the values of $\mu$, the FM correlation is decreased, and at $\mu\sim-0.2t$, the correlation vanishes.
For the A-B case, we find the AFM correlation is the strongest at $\mu=0$, and the values of the AFM correlation is decreased as we switch the chemical potential away from the zero-energy. The behavior of the spin-spin correlation in bulk graphene is completely different from that in zigzag edge of graphene ribbon. As we switch down $\mu$, there exists an interval of $\mu$ that the correlation increases for decreasing $\mu$. In zigzag graphene ribbon, due to the zero-energy states, the local moment formation at half-filling is strongly quenched, while in bulk graphene, the vanishing DOS at the Dirac cone gives rise to a well developed local moment.

\section{Summary} \label{zgnrconclusion}
We examine the indirect coupling between two magnetic impurities located on the zigzag edge of graphene ribbon, and compare the results with those obtained in bulk graphene sheet.
Due to the zero-energy localized states on the zigzag edge, the local moment formation is strongly quenched. Interestingly, when the two impurities are located on the zigzag edge, we find an enhancement in AFM correlation with decreasing $\mu$, which dose not exist in bulk graphene. The spin-spin correlation between two edge sites (A-A) is much enhanced when comparing with its bulk graphene counterpart, indicating the increase in indirect coupling caused by the zero-energy localized states.

\section{Acknowledgement}
The authors thank Zhongbing Huang and Tianxing Ma for helpful discussions. This work was supported by the Research Grants Council of Hong Kong (402310, HKUST3/CRF/09). F. M. Hu was supported by Academy of Finland through its Center of Excellence (2012-2017) program. We acknowledge the CPU time from CUHK in Hong Kong and CSC-IT Center for Science Ltd in Finland.

%\input{REF}
%\bibliographystyle{plain}  Entries are ordered alphabetically;
%\bibliographystyle{unsrt}  %Entries are not ordered alphabetically, but in the order they are first referenced.X
%\bibliographystyle{abbrv}  The bibliography looks the same as for plain style except that first names and names of journals and months are abbreviated;
%\bibliographystyle{phaip}
%\clearpage
%\appendix

\end{document}